\documentclass{ws-ijmpd}
\usepackage[dvipdfm]{hyperref}

\newcommand{\minn}{_{\rm min}}

\newcommand{\obsmaxx}{_{\rm obs,max}}
\newcommand{\Piobsmaxx}{_{\rm\Pi obs,max}}

\newcommand{\jet}{_{\rm jet}}
\newcommand{\blob}{_{\rm blob}}
\newcommand{\bend}{_{\rm bend}}

\begin{document}

\markboth{Krzysztof Nalewajko}
{Polarization Swings from Curved Trajectories}

%
\catchline{}{}{}{}{}
%

\title{POLARIZATION SWINGS FROM CURVED TRAJECTORIES\\OF THE EMITTING REGIONS}

\author{KRZYSZTOF NALEWAJKO}

\address{Nicolaus Copernicus Astronomical Center\\
Bartycka 18, 00-716 Warsaw, Poland\\
knalew@camk.edu.pl}

\maketitle

\begin{history}
\received{Day Month Year}
\revised{Day Month Year}
\comby{Managing Editor}
\end{history}

\begin{abstract}
We present a model of polarization swings in blazars from axially symmetric blobs propagating on curved trajectories. If the minimum inclination of the velocity vector to the line of sight is smaller than $\Gamma^{-1}$, the polarization angle maximum rotation rate is simultaneous with the polarization degree minimum and a spike in the total flux. By measuring the maximum rotation rate and the moment of the polarization maximum, it is possible to estimate the distance covered by the blob and thus its approximate position. We apply this model to the recent polarization event in blazar 3C~279.
\end{abstract}

\keywords{Polarization; galaxies: jets; quasars: individual (3C~279).}

\section{Introduction}

Radio to optical emission of relativistic jets in Active Galactic Nuclei (AGNs) is linearly polarized and this fact gives the strongest support for its synchrotron origin. Both polarization degree (PD) and polarization angle (PA) are strongly variable. Most of the time their behaviour is rather chaotic, but occasionally they show coherent events. Two kinds of PA variations have been observed: systematic rotations of amplitude much larger than $180^\circ$ on timescales of months to years ({\it e.~g.} in PKS 0721-115\cite{1981ApJ...248L...5A} and PKS 0521-365\cite{1993A&A...269...77L}) and fast swings of amplitude close to $180^\circ$ ({\it e.~g.} in S5 0917+624\cite{1989A&A...226L...1Q} and BL Lac\cite{2008Natur.452..966M}).

Theoretical efforts to explain the PA rotation in AGN jets began with a model of accelerating clumps of matter\cite{1979ApJ...232...34B}.  A net transverse magnetic field component is required to break the axial symmetry of the emitting region, so that the resulting PA rotation is smooth and gradual. It was later shown that other kinematic effects causing variations of the viewing angle in comoving frame (jet bending or reorientation) could trigger a PA swing\cite{1982ApJ...260..855B}. These models could explain PA swings of amplitude up to $180^\circ$. For larger PA rotations, models involving helical\cite{1985ApJ...289..188K} or stochastic\cite{1985ApJ...290..627J} magnetic fields have been proposed.

In all these early works some kind of non-axisymmetric internal jet structure was assumed. Here we show, that it does not need to be the case. A symmetric emitting region propagating on a bent trajectory, not necessarily helical, can produce a gradual PA swing. This is arguably the simplest way to reproduce several features in the simultaneous behaviour of PA and PD in some observational datasets with a minimum number of parameters.

\section{Planar bent trajectory}

\begin{figure}
\centerline{
\psfig{file=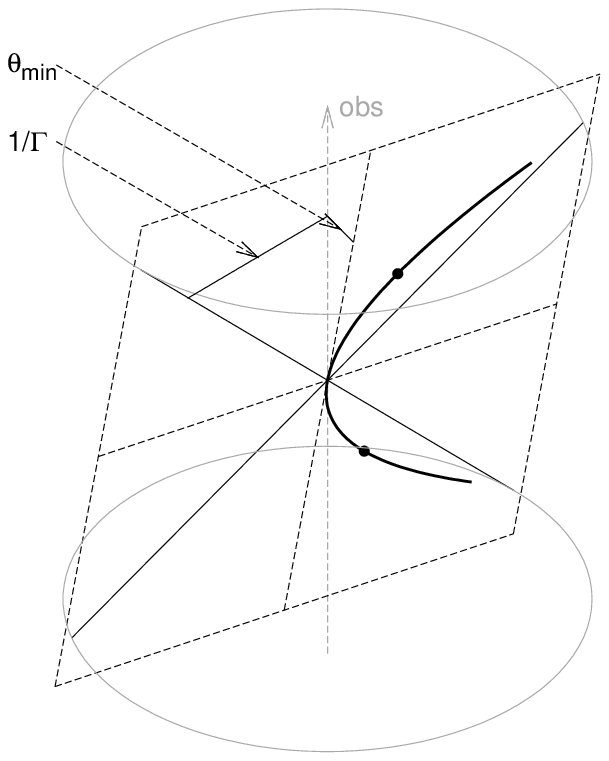,width=0.72\textwidth}
\psfig{file=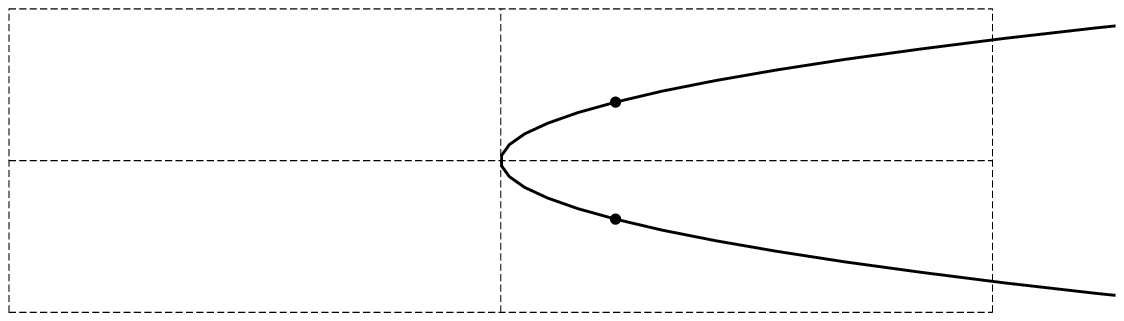,height=0.28\textwidth,angle=90}
}
\caption{The simplest case of blob trajectory that would produce a smooth PA swing of amplitude $180^\circ$. Schematically shown is its orientation with respect to the line of sight (\emph{left}) and the shape projected on the sky (\emph{right}). The trajectory is confined to a plane inclined to the line of sight at angle $\theta\minn$. The 'Doppler cone' of opening angle $\sim 1/\Gamma\jet$ is indicated. Black points on the trajectory mark the positions, at which the velocity vector is aligned with the Doppler cone, {\it i.~e.} its inclination to the observer is $1/\Gamma\jet$ in the external frame and $90^\circ$ in the co-moving frame. At these points PD reaches its maxima.}
\label{f1}
\end{figure}

Let us assume that for the time the PA is observed, the jet emission is dominated by a single compact emitting region ('blob') filled with magnetic fields axially symmetric with respect to the velocity direction. It is known, that in such a case the polarization electric vector would be either perpendicular or parallel to the projected symmetry axis\cite{1990ApJ...350..536C}, so it would be determined by the blob trajectory. To obtain a smooth PA rotation, the trajectory has to be curved and confined (at least locally) to a plane tilted to the line of sight by angle $\theta\minn$ (see Fig.~\ref{f1}). The simplest possible case is that of planar curve of constant curvature radius $R$ (circle) and constant blob velocity $\beta\blob$. When projected on the plane of the sky, it would appear as a curve bent by an angle equal to the observed PA rotation amplitude, {\it i.~e.} $\sim 180^\circ$. Interestingly, such large jet bends have been observed with VLBI on kpc scales in PKS 1510-089\cite{2002ApJ...580..742H} and PKS 2136+141\cite{2006ApJ...647..172S}.

The PA rotation rate would achieve its maximum $(d\chi/dt)\obsmaxx$ at moment $t_0$, when velocity vector inclination to the line of sight $\theta$ is minimal and equal to $\theta\minn$. For perpendicular magnetic fields, PD is highest for $\theta'=\pi/2$ (as measured in blob co-moving frame), or $\sin\theta=\Gamma\jet^{-1}$ (in external frame), where $\Gamma\jet$ is the Lorentz factor of the jet flow. PD would drop to 0 for $\theta'=0=\theta$, thus for $\theta\minn<\Gamma\jet^{-1}$ we expect a minimum of PD exactly at $t_0$. The behaviour of PD should be a symmetric function of time with two maxima at $t_0\pm \Delta t\Piobsmaxx$.

By measuring $(d\chi/dt)\obsmaxx$ and $\Delta t\Piobsmaxx$ from the observational data, we can put the constraints on the jet parameters. Taking into account the light-travel effects, we obtain:
\begin{eqnarray}
\left.\frac{d\chi}{dt}\right|\obsmaxx &=& \frac{\Omega}{\left(1-\beta\blob\cos\theta_{\rm min}\right)\sin\theta_{\rm min}}\,,\\
\Delta t\Piobsmaxx &=& \frac{1}{\Omega}\left[\arccos\left(\frac{\beta\jet}{\cos\theta_{\rm min}}\right)-\beta\blob\sqrt{\cos^2\theta_{\rm min}-\beta\jet^2}\right]\,,
\end{eqnarray}
where $\Omega=\beta\blob c/R$ is the angular velocity of the blob velocity direction. By assuming $\Gamma\jet$ and $\Gamma\blob$, we can calculate $\theta\minn$ and $R$ and find the distance covered by the blob between the PD minimum and maximum:
\begin{equation}
\label{eq_deltar}
\Delta r\blob=R\times\arccos\left(\frac{\beta\jet}{\cos\theta\minn}\right)\,.
\end{equation}
This gives us the lower bound for the position of the emitting region $r\blob \gtrsim \Delta r\blob$.

\section{Application to a polarization swing in 3C~279}

\begin{figure}
\centerline{
\psfig{file=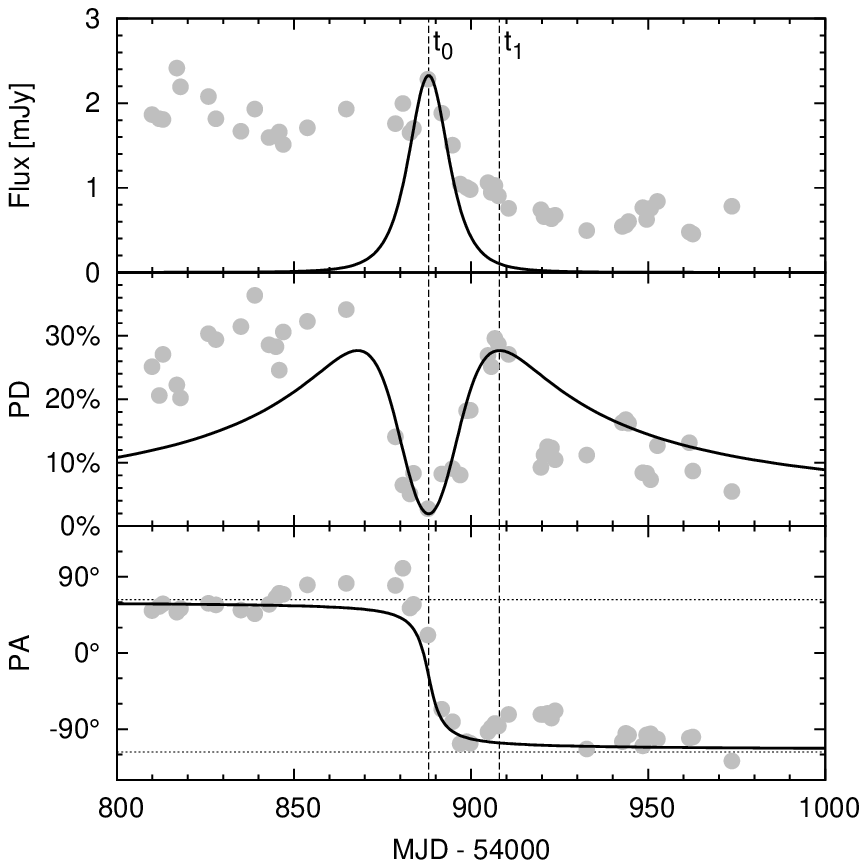,width=0.6\textwidth}
\psfig{file=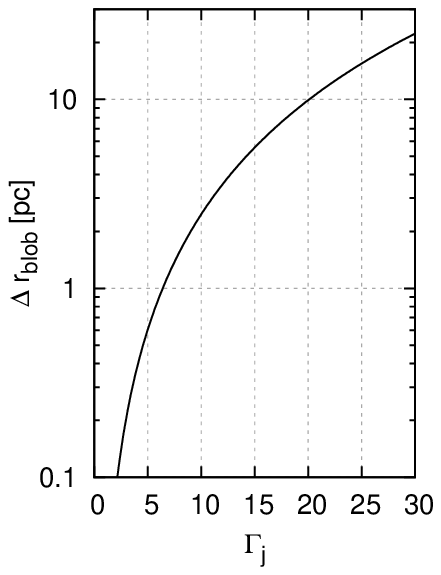,width=0.4\textwidth}
}
\caption{\emph{Left:} V-band flux and polarimetric data for 3C~279 as observed with KANATA telescope. A model for polarization swing from the bent jet for observed maximum PA rotation rate $(d\chi/dt)\obsmaxx=20^\circ/{\rm d}$, time elapsed between PD minimum and maximum $\Delta t\Piobsmaxx=t_1-t_0=20\;{\rm d}$ and bulk Lorentz factor $\Gamma\jet=\Gamma\blob=15$ is shown with a \emph{solid line}. \emph{Vertical dashed lines} mark the moment of PD minimum $t_0={\rm MJD}\;54888$ and the moment of PD maximum $t_1={\rm MJD}\;54908$. \emph{Horizontal dotted lines} mark asymptotic PA values of $-117^\circ$ and $63^\circ$. \emph{Right:} distance travelled by the blob between $t_0$ and $t_1$, as calculated from Eq. (\ref{eq_deltar}) for $\Gamma\jet=\Gamma\blob$.}
\label{f3}
\end{figure}

Recently presented\cite{2009Masaaki} polarimetric observations of blazar 3C~279 with KANATA and KVA telescopes show PA swing with reported amplitude of $\sim 208^\circ$ coinciding with a significant PD decrease to $3\%$ (see Fig. \ref{f3}). The PD minimum coincides well with the midpoint of PA rotation and a peak in the total flux at $t_0={\rm MJD}\;54888$. Both before and after the swing PA shows little variability for a couple of months and is consistent with VLBI jet orientation $\chi\jet\sim 60^\circ$\cite{2008ApJ...689...79C}. $t_0$ also marks a significant decrease in the optical and $\gamma$-ray flux. After $t_0$ PD rises to achieve a maximum of $25\%$ at $t_1={\rm MJD}\;54908$, during which both PA and the total optical flux are very stable. This behaviour was not parallelled before the swing, as PD was higher ($\sim 30\%$) and no clear peak can be seen.

We propose a scenario, in which there is a major bending in the jet at some $r\bend$. The high flux state was dominated by emission from the inner jet ($r<r\bend$) and the flux state transition was caused by a sudden decrease in the central engine activity. Then the last portion of strongly emitting plasma travelled down the bent outer jet, producing a coherent PA swing with the successive PD maximum. We can thus use the bent-trajectory model to estimate $r\bend$. We adopt $\Delta t\Piobsmaxx=t_1-t_0=20\;{\rm d}$ and $(d\chi/dt)\obsmaxx=20^\circ/{\rm d}$. Assuming the Lorentz factor $\Gamma\blob\simeq\Gamma\jet=15$, consistent with VLBI measurements of the superluminal motions\cite{2005AJ....130.1418J}, one can uniquely determine the orientation ($\theta\minn\simeq0.7^\circ$) and curvature radius ($R\simeq 85\;{\rm pc}$) of the blob trajectory and calculate the Stokes parameters and the Doppler factor as a function of observed time. We plot calculated polarization parameters and flux in Fig. \ref{f3}. PA has been shifted by a constant value to obtain asymptotic values of $-117^\circ$ and $63^\circ$. Note that the observational data matches the simulated curve with exception of small departures just before and after the swing. The observed PD has been scaled down by a constant factor to match the maximum value to the observed peak of $\sim 28\%$. In this scaling the minimal value of $\sim 2\%$ at $t_0$ roughly agrees with the observations. The curve shown on the flux plot is the Doppler boosting factor $\mathcal{D}^{3+\alpha}$, where $\alpha=1.75$ is the optical spectral index ($F_\nu\propto\nu^{-\alpha}$). It has been normalized by a constant factor to match the observed flux peak. We note, that if $\mathcal{D}\simeq\Gamma\jet$ at $t_1$ and $\mathcal{D}<(1+\beta\jet)\Gamma\jet$ at $t_0$, the ratio of the Doppler factors is close to $2$ and the ratio of the fluxes at $t_0$ and $t_1$ is about $25$. Thus, the flux component from a single blob takes a form of a short spike. It matches well the observed feature at $t_0$, but cannot explain the long-term emission trend. We conclude that a rather small fraction of the observed flux has been emitted by a single blob.

Using Eq. (\ref{eq_deltar}), we calculate the distance $\Delta r\blob$ covered by the blob between the PD minimum at $t_0$ and the PD maximum at $t_1$ (see Fig. \ref{f3}). Since $\theta\minn\ll \Gamma\jet^{-1}$, it scales approximately like $\Delta r\blob \sim c\,\Delta t\Piobsmaxx\Gamma\blob^2$ . For $\Gamma\jet=\Gamma\blob=15$, we expect the jet bend to be located at $r\bend\gtrsim\Delta r\blob\sim 5\;{\rm pc}$.

\section{Discussion}

Our model is most suitable for PA swings of $\sim 180^\circ$ with simultaneous PD minimum and brightness spike, and with characteristic fast rise and slow decay of PD. We note that reported amplitude $208^\circ$ of the polarization rotation in 3C~279 is based on the values measured immediately before the event, when there was a small rotation in the other direction. We also predict an apparent $\sim 180^\circ$ twist in a pc-scale jet, that in general could be verified with high-resolution VLBI observations. However, in the case of 3C~279 this is not plausible, since there is little variability in the mm wavelenghts and the optical/$\gamma$-ray flare is produced within the mm photosphere.

Larger observed amplitudes of PA rotation, {\it e.~g.} $\sim 240^\circ$ in BL Lac\cite{2008Natur.452..966M}, could be explained by helical trajectories. If the emission source is located in the Poynting-flux-dominated region, they would be determined by conservation of angular momentum inherited from the central engine. However, they could also arise in the interactions of the jet with non-uniform and variable environment on several-pc scales. There are two observational facts supporting this scenario in the case of 3C~279. First, the kpc-scale jet of continuously changes its positional angle and the apparent velocity of the radio knots on timescale of years\cite{2008ApJ...689...79C}. Second, a PA rotation in the other direction has been observed in earlier epoch\cite{2008A&A...492..389L}, this challenges the models predicting fixed helicity sign.

If the emitting region is large enough with respect to the curvature radius, time delays between the arrivals of photons emitted from different positions across the emitting volume would affect the variation of polarization parameters. This would provide a practical method to estimate the radius of the emitting region. We could also be able to place a constraint on the magnetic field distribution across the blob extension, in particular distinguish between globally ordered (toroidal) and disordered (chaotic) magnetic fields.

Long and intensive polarimetric monitoring of bright blazars is very important, since polarization swings are fast and rare events, that cannot be predicted in advance. A good coverage is essential for verifying competing theoretical models. In the case of 3C~279 an important question is, whether the observed polarization swing was related to the transition from high to low $\gamma$-ray flux, or was it purely coincidental. We will only know this, when we observe another polarization swing in similar circumstances.

\section*{Acknowledgments}

The author thanks Marek Sikora, Greg Madejski, Masaaki Hayashida, {\L}ukasz Stawarz and Rafa{\l} Moderski for discussions and support, and acknowledges the hospitality of the Kavli Institute for Particle Astrophysics and Cosmology at Stanford Linear Accelerator Center at the time of preparing this contribution. This work was partially supported by the Polish Ministry of Science and Higher Education grants N N203 301635 and N N203 386337, and the Polish Astroparticle Network 621/E-78/BWSN-0068/2008.

\end{document}